\documentclass{svjour3} 
\usepackage{graphicx}
\usepackage[fleqn]{amsmath}

\usepackage{multirow}
\usepackage{enumitem}

\usepackage[fleqn]{amsmath}
\usepackage{array}

\usepackage{algorithm}
\usepackage{algorithmic}

\def\makeheadbox{{%
		\hbox to0pt{\vbox{\baselineskip=10dd\hrule\hbox
				to\hsize{\vrule\kern3pt\vbox{\kern3pt
						\hbox{\bfseries Knowl Inf Syst}
						\hbox{doi: 10.1007/s10115-017-1134-1}
						\kern3pt}\hfil\kern3pt\vrule}\hrule}%
			\hss}}}

\begin{document}

\newcommand{\theHalgorithm}{\arabic{algorithm}}

\title{Sentiment Analysis of Financial News Articles using Performance Indicators} 
\author{Srikumar Krishnamoorthy}

\institute{ 
	{This is an accepted version of the article published in Knowledge and Information Systems Journal. The final authenticated version is available online at: https://doi.org/10.1007/s10115-017-1134-1}
	\vspace*{1.2em} \\	     Indian Institute of Management Ahmedabad, India
              Tel.: +91-79-66324834\\
              \email{srikumark@iima.ac.in}           
}
\date{Received: 24 Sep 2016 / Revised: 30 May 2017 / Accepted: 29 Oct 2017}

\maketitle

\begin{abstract}
Mining financial text documents and understanding the sentiments of individual investors, institutions and markets is an important and challenging problem in the literature. Current approaches to mine sentiments from financial texts largely rely on domain specific dictionaries. However, dictionary based methods often fail to accurately predict the polarity of financial texts. This paper aims to improve the state-of-the-art and introduces a novel sentiment analysis approach that employs the concept of financial and non-financial performance indicators. It presents an association rule mining based hierarchical sentiment classifier model to predict the polarity of financial texts as positive, neutral or negative. The performance of the proposed model is evaluated on a benchmark financial dataset. The model is also compared against other state-of-the-art dictionary and machine learning based approaches and the results are found to be quite promising. The novel use of performance indicators for financial sentiment analysis offers interesting and useful insights.  \\

\end{abstract}

\section{Introduction}
Sentiment analysis and opinion mining \cite{Pang2002,Turney2002} has received significant attention in the literature due to its wide applicability in business, management and social science disciplines. It has been effectively applied in domains such as movies \cite{Pang2002,Turney2002}, product reviews \cite{Blitzer2007}, travel reviews \cite{Dang2010,Turney2002} and finance \cite{Antweiler2004,Huang2014,Loughran2015,Loughran2016,Tetlock2007,Tetlock2008}. 
\par 
Financial sentiment analysis is considered to be an important and challenging problem in the literature \cite{Loughran2015,Loughran2016}. Current approaches to financial sentiment analysis utilize generic dictionary \cite{Tetlock2007,Tetlock2008}, domain specific dictionary \cite{Ferguson2014,li2014effect,li2014news,Loughran2011,Malo2014} or statistical/machine learning methods \cite{Antweiler2004,Huang2014,Li2010,Malo2014,OHare2009,VanDeKauter2015} to determine polarity in financial texts. Some of the common dictionaries used in the financial sentiment analysis literature include Harvard GI (HGI) \cite{Stone1962}, MPQA \cite{Wiebe2005}, Sentiwordnet \cite{esuli2006sentiwordnet}, SenticNet \cite{cambria2014senticnet}, SentiStrength2 \cite{thelwall2012sentiment}, LM \cite{Loughran2011}, and Financial Polarity Lexicon (FPL) \cite{Malo2014}. LM \& FPL are finance-specific dictionaries used in the recent literature. Other dictionaries (such as HGI \cite{Stone1962}, MPQA \cite{Wiebe2005}, SenticNet \cite{cambria2014senticnet}) are generic in nature and are likely to mis-classify common financial words. For example, a detailed study of HGI \cite{Stone1962} dictionary by \cite{Loughran2011} shows that more than 75\% of the negative words used in HGI are non-negative in a financial context. Furthermore, recent research studies demonstrate superior results while using domain specific dictionary over general dictionary \cite{Huang2014,li2014news,Loughran2015,Malo2014}. 
\par 
In this paper, we aim to improve the state-of-the-art in domain specific dictionary based financial sentiment analysis. We motivate the need for the current study with the help of a set of illustrative examples. Let us consider the following financial text sentences extracted from financial phrase-bank dataset \cite{Malo2014}. 
 
\begin{enumerate}[leftmargin=1cm]
\item	Halonen's office acknowledged receiving the letter but \underline{declined} comment. 
\item	Financial details were not \underline{disclosed}.
\item	The serial bond is part of the plan to \underline{refinance} the short-term credit facility.
\item	Aspo's \underline{strong} company brands – ESL Shipping, Leipurin, Telko and Kaukomarkkinat – aim to be the market leaders in their sectors.
\item	DnB Nord of Norway is the most likely Nordic buyer for Citadele, while Nordea would be a \underline{good} strategic fit, according to published documents.
\item	We are also \underline{pleased} to welcome the new employees.
\end{enumerate}

\par 
In the LM dictionary \cite{Loughran2011}, declined, disclosed and refinance are negative words; similarly, strong, good and pleased are positive words. Therefore, each of the above sentences will be classified as positive or negative sentences by methods that use LM or other related domain dictionaries. However, it is quite evident that all of the above sentences are neutral statements from an investor/analyst perspective. 
\par 
A few research studies have attempted to go beyond just polarity based dictionaries and utilized financial entities (custom words \cite{Malo2014} or noun phrases \cite{li2014effect} or named entities \cite{schumaker2009textual}) to improve the quality of polarity detection. But, the use of financial entities often generate a lot of false positives and false negatives. Let us consider the following sentences that contain at least one financial entity. 

\begin{enumerate}[leftmargin=1cm]
\item	The company's board of directors has proposed a \textit{dividend} of EUR 0.12 per share.
\item	Amanda \textit{Capital} has \textit{investments} in 22 private \textit{equity funds} and in over 200 unquoted companies mainly in Europe.
\item	A corresponding \underline{increase} of 85,432.50 euros in Ahlstrom's \textit{share capital} has been entered in the Trade Register today.
\end{enumerate}

\par 
All of the above sentences are neutral statements, though they contain financial terms/entities (\textit{italized} words in text). Sentence 3 also indicates increase of a financial entity (share capital). Hence, the sentence is likely to be considered as positive by systems that use a combination of LM dictionary and financial entities (or noun phrases/named entities) \cite{li2014effect,Malo2014}.  
\par 
The following sentences contain financial or non-financial terms and directionality related words (e.g. raise, fell, lower, increase and so on). 
\begin{enumerate}[leftmargin=1cm]
\item	\textit{Turnover} \underline{rose} to EUR 21mn from EUR 17mn.
\item	The \textit{EPS} outlook was \underline{increased} by 5.6 
\item	Unit \textit{costs} for flight operations \underline{fell} by 6.4 percent.
\item	The company intends to \underline{raise} \textit{production capacity} in 2006.
\item	Rapala said it estimates it will make \underline{savings} of 1-2 million euros a year by centralising its French \textit{operations} at one site.
\item	\textit{Demand} for fireplace products was \underline{lower} than expected, especially in Germany.
\end{enumerate}
\par 
Sentences 4, 5 and 6 do not use financial entities. The systems that use only financial entities are likely to treat them as neutral sentences. An investor/analyst, however, is likely to consider the above sentences as polarized ones (positive or negative). A closer examination of these sentences reveals that they: (1) use non-financial indicators like demand, production capacity, and operations, and (2) refer to improvement or decline of the non-financial indicators. 
\par 
From the foregoing discussions, it is evident that the use of domain specific dictionary (on a stand-alone basis or in conjunction with financial entities) is inadequate to accurately predict the polarity in financial texts. 
\par 
It is quite intuitive to understand that an investor or a financial analyst is likely to analyze the performance of a company in terms of both financial and non-financial performance indicators. An investor/analyst often looks for performance indicators or measures while assessing polarity in financial texts and making investment decisions. Therefore, we posit that the use of financial and non-financial performance indicators can improve the quality of financial sentiment analysis. 
\par
Our main objective in this research work is to examine the use of performance indicators (lagging/leading financial and non-financial indicators) to predict polarity in financial texts. We conjecture that the use of performance indicators \cite{Kaplan1996} is likely to be more meaningful and closely aligned with the way humans interpret financial texts while making investment decisions. We present a new approach to perform financial sentiment analysis based on performance indicators and conduct rigorous experiments to assess its usefulness. Such an approach, to the best of our knowledge, has not been explored in the literature. 
\par 
This paper primarily makes two key contributions to the literature. First, the paper introduces the use of performance indicators to predict polarity in financial texts. Second, the paper presents a hierarchical sentiment classifier model based on the concept of association rule mining \cite{Agrawal1993}. The primary benefit of using association rule mining for polarity prediction is that the generated rules can be used to easily explain predictions, unlike more complex discriminative models (like SVM, Neural networks) that are black-boxes. The performance of the proposed sentiment classifier model is assessed on a benchmark financial dataset. The model is also compared against other state-of-the-art methods to demonstrate its usefulness. 
\par 
The rest of the paper is organized as follows: In the next section, the related work in the literature is described. Subsequently, in section 3, the paper outlines the proposed method in detail. The experimental evaluation of the method and a discussion of the key findings are presented in section 4. The paper concludes with a summary and directions for future research work.

\section{Related Literature}
Financial sentiment analysis approaches in the literature can be broadly categorized as (a) generic dictionary based methods, (b) domain specific dictionary based methods, and (c) statistical or machine learning based methods. Generic dictionaries such as Harvard GI \cite{Stone1962} was used in some of the early works in financial sentiment analysis \cite{Tetlock2007,Tetlock2008}. The use of generic dictionaries, however, lead to misclassification of common words in financial text and impacts the performance of sentiment prediction \cite{Loughran2011} or stock price movements \cite{li2014effect,li2014news}. Recent works in the literature \cite{Ferguson2014,li2016tensor,li2014effect,li2014news,Malo2014} predominantly use domain specific dictionary such as LM dictionary \cite{Loughran2011}, and FPL \cite{Malo2014}. A detailed review of different dictionaries used in the literature can be found in \cite{Loughran2015}. Recent surveys on sentiment analysis in finance can be found in \cite{Kearney2014,Loughran2016}.
\par 

Statistical or machine learning based methods \cite{Antweiler2004,Huang2014,Li2010,Malo2014,OHare2009,VanDeKauter2015} use bag-of-words or n-grams as features and apply generative or discriminative classifier models for predicting sentiments. Internet message postings were used by \cite{Antweiler2004} to classify financial text as buy, hold or sell. The classifier results were then aggregated for a pre-defined time period and bullishness \& agreement indices were computed. The authors also conducted a study on relationship between the computed indices and financial measures such as stock returns and market volatility. The experimental results show that stock messages help predict market volatility. 
\par 
A naive bayes classifier model is used in \cite{Huang2014} to classify sentences as positive, negative or neutral. The predicted sentence level opinions are aggregated to derive report level opinions. The authors show that their method outperforms both generic and domain specific dictionary methods. A naive bayes classifier model based on bag-of-words features is trained in \cite{Li2010} to predict the tone (as positive, neutral, negative or uncertain) of forward looking statements in corporate filings. The author presents evidence that dictionary based methods (both generic and domain specific) are unsuitable for analyzing tone of corporate filings. 

\begin{table}[!htb]
\caption{Comparison of financial sentiment analysis literature}
\label{table:litSummary}
\begin{tabular}{c|c|c|c|c|c}
\hline\noalign{\smallskip}
Literature & \parbox[t]{1cm}{Nature \\of text} & Dictionary & Features & Method & Objective \\
\noalign{\smallskip}
\hline 
\noalign{\smallskip}
Tetlock \cite{Tetlock2007} & News & HGI & HGI words & Regression & \parbox[t]{1.9cm}{\centering Stock \\returns} \\ 
{} & {}& {}& {}& {}& {} \\
Tetlock \cite{Tetlock2008} & News & HGI & HGI words & Regression & \parbox[t]{1.9cm}{\centering Earnings, \\ Returns} \\
{} & {}& {}& {}& {}& {} \\
Loughran \cite{Loughran2011} & 10-K & LM & LM words & Regression & \parbox[t]{1.9cm}{\centering Returns} \\
{} & {}& {}& {}& {}& {} \\
Antweiler \cite{Antweiler2004} & \parbox[t]{1cm}{Internet \\messages} & - & BoW & \parbox[t]{1.5cm}{\centering NB \\ Regression} & \parbox[t]{1.9cm}{\centering Volatility, \\ Returns} \\  
{} & {}& {}& {}& {}& {} \\
Huang \cite{Huang2014} & \parbox[t]{1cm}{Analyst \\reports} & \parbox[t]{1cm}{\centering HGI,LM\\LIWC\\Diction} & BoW & \parbox[t]{1.9cm}{\centering NB \\ Regression} & \parbox[t]{1.5cm}{\centering Polarity, \\ Earnings} \\
Li \cite{Li2010} & \parbox[t]{1cm}{\centering 10-K\\10-Q} & - & BoW & NB & \parbox[t]{1.9cm}{\centering Polarity, \\ Info. Content} \\
{} & {}& {}& {}& {}& {} \\
O'Hare \cite{OHare2009} & Blogs & - & \parbox[t]{1.5cm}{\centering Topic terms, \\Words near topics} & NB,SVM & \parbox[t]{1.9cm}{\centering Polarity \\prediction} \\
Marjan \cite{VanDeKauter2015} & News & Dutch & \parbox[t]{1.5cm}{\centering Entities \\Polarity expression} & Unsupervised & \parbox[t]{1.9cm}{\centering Polarity \\detection} \\  
Li \cite{li2016tensor} & \parbox[t]{1cm}{\centering News, \\Social media} & - & \parbox[t]{1.5cm}{\centering Nouns, Sentiment words} & \parbox[t]{1.5cm}{\centering Tensor \\regression} & \parbox[t]{1.9cm}{\centering Stock \\trends} \\
Li \cite{li2014news} &  News & HGI,LM & \parbox[t]{1.5cm}{\centering Dictionary \\words} & SVM & \parbox[t]{1.9cm}{\centering Stock \\returns} \\
{} & {}& {}& {}& {}& {} \\
Li \cite{li2014effect} &  \parbox[t]{1cm}{\centering News, \\Internet messages} & \parbox[t]{1.5cm}{\centering HGI,\\LM} & \parbox[t]{1.5cm}{\centering Entities,\\Polarity words} & SVR & \parbox[t]{1.9cm}{\centering Stock \\trends} \\
{} & {}& {}& {}& {}& {} \\
Mo \cite{mo2016news} & News & SWN & \parbox[t]{1.5cm}{\centering SWN \\words} & Regression & \parbox[t]{1.9cm}{\centering Stock \\returns} \\
{} & {}& {}& {}& {}& {} \\
RPS \cite{schumaker2009textual} & News & - & Nouns & SVR & \parbox[t]{1.9cm}{\centering Stock \\trends} \\
{} & {}& {}& {}& {}& {} \\
Malo \cite{Malo2014} & News & \parbox[t]{1.2cm}{\centering FE,MPQA\\HGI,LM\\DI} & \parbox[t]{1.5cm}{\centering Entities, \\Phrase structure} & SVM & \parbox[t]{1.9cm}{\centering Polarity \\prediction} \\
{} & {}& {}& {}& {}& {} \\
Our work & News & \parbox[t]{1cm}{\centering PI, LM\\DI} & PI tags & ARM & \parbox[t]{1.9cm}{\centering Polarity \\prediction} \\ 
{} & {}& {}& {}& {}& {} \\
\hline
\end{tabular}
BoW - Bag of words; FE - Financial entity; DI - Directionality; SWN - Sentiwordnet \\
PI - Performance indicator words; LIWC - Linguistic Inquiry and Word Count 
\end{table}

\par 
Naive bayes and support vector machine methods were used to predict sentiment of bloggers towards companies and their stocks \cite{OHare2009}. The sentiment prediction was done using selected topic terms. Their work analyzes sentiment at the topic level and not at the sentence level as in \cite{Li2010,Malo2014}. Other recent works in the literature that perform sentiment analysis include \cite{li2016tensor,li2014effect,li2014news,mo2016news,schumaker2009textual}.
These works utilize sentiment dictionaries and extract sentiments from text and predict stock price movements or market returns. The proposed paper is distinct from such works as the focus of the current paper is primarily on predicting the polarity of news articles as positive, neutral or negative. A comparative analysis of the financial sentiment analysis works in the literature is presented in Table~\ref{table:litSummary}. 
\par 
The polarity sequence model proposed in \cite{Malo2014} is an extension of \cite{Moilanen2010}. The authors propose an LPS model and utilize both generic dictionary, MPQA \cite{Wiebe2005} and domain specific LM dictionary \cite{Loughran2011}. The authors also enrich the finance lexicon by including (a) financial concept, (b) directional verbs such as increase, decrease, and (c) polarity for the interaction between financial concept and directional verb e.g. cost-increase is pre-labelled in the dictionary as negative; profit-increase is pre-labelled as positive. The authors also contribute to the literature by making an annotated financial sentiment corpus publicly available.
\par 
The LPS model proposed in \cite{Malo2014} primarily works in three phases. In the first phase, entities are extracted along with their semantic orientations from the given text sentences. Several heuristics are applied to extract different entities (financial and general) and semantic orientations from financial text sentences. Subsequently, in the second phase, the model applies a phrase structure projection to project the extracted entities into a sequence in $l^{2}$ space. Finally, in the third phase, a multi-label classifier is applied on the generated $l^{2}$ sequence. For multi-label classification, support vector machines (SVM) with one-against-one strategy was used. 
\par 
The proposed work is distinct from existing works in the literature on the following aspects: (1) It presents a financial text polarity prediction method based on financial and non-financial performance indicators. This is different from the works of \cite{Malo2014} that primarily use financial concepts and are likely to generate a lot of false positives and false negatives. Our experimental results corroborates this claim and show the usefulness of the proposed method. (2) Polarity for the interaction between financial concept and directional verb is not pre-defined as in \cite{Malo2014}. (3) Presents an association rule \cite{Agrawal1993} based hierarchical classifier to predict sentiment. To the best of our knowledge, this is the first work that uses association rule mining based method for financial sentiment analysis. A few recent works in non-financial domain have explored the use of a non-hierarchical associative classifier for sentiment analysis of webreviews \cite{man2014investigating} and product reviews \cite{yang2010understanding}. 
\par 
 We demonstrate the usefulness of the proposed approach over other state-of-the-art methods through rigorous experimental evaluation.

\section{Hierarchical Sentiment Classifier (HSC)}

This section describes the proposed method for financial sentiment analysis. The method broadly consists of the following four key aspects.
\begin{enumerate}[leftmargin=1cm]
\item	\textbf{Domain specific lexicon} The proposed method uses a standard domain specific dictionary, LM dictionary \cite{Loughran2011}. In addition, the lexicon defines words related to performance indicators and directionality. 
\item	\textbf{Text tagging using lexicon} The given financial text is parsed and tagged by looking up words in the lexicon. 
\item	\textbf{Polarity classifier model} The tagged financial text is used to build a hierarchical classifier model. The classifier model utilizes the concept of association rule mining. 
\item   \textbf{Predict sentiment} The polarity classifier model (association rules) is used to make polarity predictions for new financial text sentences.  
\end{enumerate}

\par 
Each of the above aspects is described in detail in the following pages. 

\subsection{Domain specific lexicon}
Three categories of words are defined in the domain specific lexicon used in this paper. The overall distribution of words in the lexicon is given in Table~\ref{table:wordDistribution}.

\begin{table}[!htb]
\caption{Distribution of words in the dictionary}
\label{table:wordDistribution}
\begin{tabular}{p{3cm} >{\centering}p{4.9cm} >{\raggedleft\arraybackslash}p{2cm}}
\hline\noalign{\smallskip}
Category & \parbox[t]{4.9cm}{Type of word (tags)} & \parbox[t]{3cm}{No. of words \\ (\% of all entries)} \\
\noalign{\smallskip}
\hline 
\noalign{\smallskip}
\parbox[t]{3cm}{Performance} & \parbox[t]{4.9cm}{Lagging Indicator words (LagInd)} & 67 (2.29\%) \\
\parbox[t]{3cm}{Indicators} & \parbox[t]{4.9cm}{Leading indicator words (LeadInd)} & 70 (2.39\%) \\
\\
Directionality & \parbox[t]{4.9cm}{Down (DOWN)} & 53 (1.81\%)  \\
\parbox[t]{3cm}{} & \parbox[t]{4.9cm}{Up (UP)} & 51 (1.74\%) \\
\\
\parbox[t]{3cm}{Finance-specific} & \parbox[t]{4.9cm}{Negative (NEG)} & 2337 (79.73\%) \\
\parbox[t]{3cm}{sentiment words} & \parbox[t]{4.9cm}{Positive (POS)} & 353 (12.04\%) \\
\hline
\end{tabular}
\end{table}

The first category of words is related to performance indicators. The performance indicator words are categorized as lagging and leading indicators. Lagging indicators reflect the results of firm’s activity e.g. improvement or decline in sales, market share, operating profit, operating cost, orders, inventory turns etc. 
\par 
The leading indicators signal future events and are precursors to the firm’s future performance. Some of the common examples of leading indicators are: \#new stores, \#employee recruitments, \#employee reductions/layoffs, \#new customers, \% increase in productivity, \% increase in production capacity, \#new contracts won or awarded, and \% increase in plant utilization. 
\par 
The specific words used in the dictionary for leading indicators include only the key terms like store, employee, customer, productivity, efficiency. The words that signify the direction of movement of these terms (i.e. increase, decrease, reduction, layoff etc.) are maintained separately as part of the directionality category that is described next. 
\par 
The second category of words defined in the lexicon is related to the directionality of leading/lagging indicators. Directionality is the word (or n-gram) describing direction of events \cite{Malo2014}. Examples include increase, improve, skyrocket, decrease, and plummet. For building the directionality related words, the words defined under specific categories (namely, rise, fall, increase, and decrease) in the Harvard GI lexicon \cite{Stone1962} were used as seed words. The words were manually reviewed to remove words that have different meaning in financial context. A few additional words that signify directionality of lagging/leading indicator terms were included e.g. layoff (\#employees laid off), terminated (\#staff terminated), and awarded (\#new contracts awarded). The final number of words under the directionality category was about 100.   
\par 
The third, and final, category of words used in the lexicon is the domain-specific lexicon, LM dictionary \cite{Loughran2011}.  The polarity bearing words defined in the LM dictionary were also used in this study.

\subsection{Text tagging using lexicon}

\begin{figure}[tbh]
\centering
\includegraphics[width=12.3cm,height=12cm]{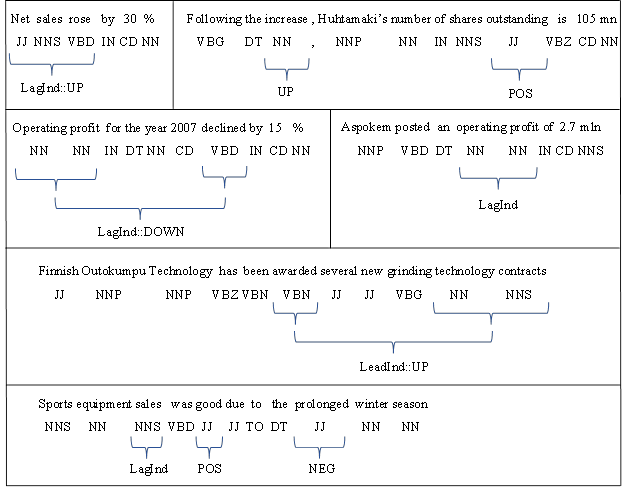}
\caption{Illustration of text sentences tagged with POS tags and lexicon related tags}
\label{figure:taggingIllustration}
\end{figure}

The input to the proposed method is a set of financial text sentences. Each financial text sentence is parsed using NLTK Parts-Of-Speech (POS) tagger \cite{bird2006nltk}. The sentences are then tagged or labeled based on the occurrence of different categories of words (i.e. performance indicators, directionality, interaction of performance indicators \& directionality, and sentiment words). The occurrence patterns of different categories of words are analyzed using several heuristic rules. Fig.~\ref{figure:taggingIllustration} provides a couple of illustrative examples of parsed and tagged text sentences. Fig.~\ref{figure:taggingIllustration} gives the raw text sentence, its corresponding POS tags and the lexicon tagged words. The lexicons used for tagging are the ones described in section 3.1. The complete grammar and the heuristic rules used for parsing and labeling financial text sentences are given in Appendix. 

The output of this step is the conversion of text sentences into tagged sequence of words. The specific tags used are: LagInd, LeadInd, UP, DOWN, POS, NEG, LagInd::UP, LagInd::DOWN, LeadInd::UP, and LeadInd::DOWN. The last four tags are the interactions between performance indicators and directionality. All other words in the original sentence are discarded. 
The above process is repeated for each and every sentence in the review collection. The resulting collection of tagged words is then used for polarity classification.

\subsection{Polarity classifier model}

The proposed method aims to classify the financial text as positive, neutral or negative. The classifier proposed in this paper uses the concept of association rule mining \cite{Agrawal1993}. Some of the common associative classifiers used in the literature include: CBA \cite{Liu1998}, CMAR \cite{Li2001}, LB \cite{meretakis1999extending} and ART \cite{berzal2004art}. The proposed approach utilizes a variation of the CMAR method for sentiment classification. The key steps of the associative classifier proposed in this paper are as follows:

\subsubsection{Prepare transaction data}
The input data required for rule mining is the set of transactions with items. The tags identified in the earlier step are used as items for rule mining. In addition, the class label of the sentence is appended to the set of items. For example, the transaction data for the illustrative example in Figure~\ref{figure:taggingIllustration} is shown in Table~\ref{table:transDB}.

\begin{table}[!htb]
\caption{A sample tagged transaction database}
\label{table:transDB}
\begin{tabular}{>{\centering}p{2cm} >{\raggedright\arraybackslash}p{4.3cm}}
\hline\noalign{\smallskip}
Transaction \# & Type of word (tags)\\
\noalign{\smallskip}
\hline 
\noalign{\smallskip}
1 & LagInd::UP, positive \\
2 & UP, POS, neutral \\
3 &	LagInd::DOWN, negative \\
4 &	LagInd, neutral \\
5 &	LeadInd::UP, positive \\
6 &	LagInd, POS, NEG, neutral \\
\hline
\end{tabular}
\end{table}

\subsubsection{Build a classifier model}
The model building involves frequent itemset mining, association rule generation, and rule ordering. 
\par 
Frequent itemset and rule mining. The frequent itemsets are mined using Apriori algorithm \cite{Agrawal1993}. Association rules are mined from the frequent itemsets with the constraint that the consequent of the rule contains one item and is a class (positive, negative, or neutral class). Antecedent of the rule can contain any number of items. That is, rules of the form: X $\rightarrow$ c, are mined from the generated frequent itemsets. A sample set of association rules mined from the transactions in Table~\ref{table:transDB} is given in Figure~\ref{figure:sampleRules}. 

\begin{figure}[tbhp]
\hspace{-4mm}
\begin{minipage}{30em}
\centering
LagInd $\rightarrow$ neutral (33.33\%, 100\%) \\
POS $\rightarrow$ neutral (33.33\%, 100\%) \\
UP, POS $\rightarrow$ neutral (16.67\%, 100\%) \\
LagInd::DOWN $\rightarrow$ negative (16.67\%, 100\%) \\
LagInd::UP $\rightarrow$ positive (16.67\%, 100\%) \\
LeadInd::UP $\rightarrow$ positive (16.67\%, 100\%) \\
UP $\rightarrow$ neutral (16.67\%, 100\%) \\
\end{minipage}
\caption{Sample Association Rules}
\label{figure:sampleRules}
\end{figure}

Rule ordering. The generated association rules are ordered according to decreasing precedence based on their confidence, support and antecedent length. This heuristic is used in the literature to give precedence to more confident rules \cite{Li2001,Liu1998} while making predictions. 
\par 
\subsection{Predict sentiment} 
The rules mined in the previous step are used as a model for predicting the polarity of new sentences. The complete pseudo code for polarity prediction is given in Algorithm~\ref{alg:pseudocode1}. 
\par 
Given a new financial text sentence, the polarity prediction works as follows: First, the new text sentence is tagged with performance indicators, directionality and finance-specific sentiment words. The tagged words are then looked up in the rule base (i.e. the set of rules generated by the polarity classifier model). It is to be noted that each rule in the rule base is of the form: Premise (one or more tag words) $\rightarrow$ Conclusion (the class label). The tagged words are matched against each and every rule in the rule base (steps 5-19 of Algorithm~\ref{alg:pseudocode1}). If a rule match is found, a matching score is assigned based on the confidence of the rule (the higher the confidence, the more important the rule). The scores are grouped based on class labels, as multiple rules with different class labels are likely to match the tagged words. The process of grouping scores based on class labels is similar to the approach followed by associative classifiers in the literature \cite{Li2001}. The above process is repeated for each and every rule in the rule base and the scores are accumulated. At the end of the iteration, the algorithm produces a score for each class label. The class with the highest score is treated as the final predicted class (step 20 of Algorithm~\ref{alg:pseudocode1}). 
\par 
Let us illustrate the above prediction procedure with the following example:
\par 
\vspace{.5em}
\hspace{1em}\vspace{.5em} \textit{Olvi expects market share to increase in the first quarter of 2010}
\par 
The above sentence after parsing and dictionary lookup will be tagged as LagInd::UP. The class label is then determined by looking up the tag in the set of mined rules. The only matching rule available in Figure~\ref{figure:sampleRules} for the tagged sentence is LagInd::UP $\rightarrow$ positive (16.67\%, 100\%). The scores are, therefore, generated only for the 'positive' class label. The most appropriate class label for the sentence is predicted as "positive". If there are rule matches across class labels, then scores for each class label are generated (refer to steps 5 to 19 of Algorithm~\ref{alg:pseudocode1}). The final class label is predicted as the one with highest score. If no rule matches are found during the search process, then the sentence is assigned a default prediction value of "neutral". 

\begin{algorithm}[!htb]
   \caption{Pseudocode for polarity prediction}
   \label{alg:pseudocode1}
{\bfseries Input:} $inputText$, financial text sentence \\ 
		\hspace*{2.7em} $ruleBase$, mined association rules\\ 
   {\bfseries Output:} $result$, the predicted class   
\begin{algorithmic}[1]
   \STATE{result=$\emptyset$} \textit{$\triangleright$ a dictionary that holds confidence values for each class}
   \STATE{tags $\leftarrow$ Parse and tag the text sentence} \#section 3.2 
   \STATE{\textit{$\triangleright$ iterate through each rule and match the tags}}
   \STATE{\textit{$\triangleright$ each rule, premise $\rightarrow$ class, contains support and confidence values}} 
   \FOR{r in $ruleBase$}  
     \STATE{\textit{$\triangleright$ check if complete set of tags match}} 
     \IF{r.premise==tags} 
       {\STATE{\textit{$\triangleright$ accumulate the score for each class in result}}}
       \STATE{result[r.class] = result[r.class] + r.confidence}   
     \ELSE
       \STATE{\textit{$\triangleright$ match each tag individually}}
       \FOR{t in tags}
         \IF{r.premise==t} 
           \STATE{\textit{$\triangleright$ one of the tags match}}
           \STATE{result[r.class] = result[r.class] + r.confidence}   
		 \ENDIF	   
	   \ENDFOR 
     \ENDIF
   \ENDFOR
   \STATE {\bfseries return} class with the highest average confidence
\end{algorithmic}
\end{algorithm}

\par 
The polarity classifier model and prediction method described above is generic and can be used for multi-class problems (positive, neutral, negative sentiments). However, in the financial sentiment analysis context, the total number of training examples in each class is likely to be highly unbalanced. That is, one is likely to encounter more number of neutral sentences than positive/negative sentences in financial texts. This class imbalance can also be observed in the annotated corpus used by researchers in the past literature. For example, the number of neutral examples was well over 53\%; and negative examples was a small fraction of about 12-14\% in the earlier works \cite{Huang2014,Malo2014}. 
\par 
In order to effectively address the class imbalance problem, we propose a hierarchical classifier, named Hierarchical Sentiment Classifier (HSC). The proposed method (HSC) first classifies a given sentence as polarized or neutral. Subsequently, the polarized sentences are classified as either positive or negative. The underlying classifier model and prediction methods used in HSC were described earlier (section 3). A comparative evaluation of the hierarchical versus the multi-class and the standard one-against-one classifier is presented in the experimental results section.

\section{Experimental Evaluation}

The publicly available financial phrase bank dataset \cite{Malo2014} is used for the experimental analysis. This dataset is a reasonably large annotated corpus of about 5000 sentences. The sentences are marked as positive, neutral or negative by the annotators. Four different datasets were prepared by the authors based on the level of agreement amongst the annotators. These datasets are labelled as DS100 (100\% agreement amongst annotators), DS75 ($>$75\% agreement), DS66 ($>$66\% agreement) and DS50 ($>$50\% agreement). The total number of examples and distribution of class labels on each of the datasets is given in Table~\ref{table:datasetCharacteristics}.

\begin{table}[!htb]
\caption{Dataset characteristics}
\label{table:datasetCharacteristics}
\begin{tabular}{>{\centering}p{2cm} >{\raggedright\arraybackslash}p{4.3cm}}
\hline\noalign{\smallskip}
Dataset & Number of sentences\\
& (\%positive, \%neutral, \%negative) \\
\noalign{\smallskip}
\hline 
\noalign{\smallskip}
DS100 & 2259 (13.4, 61.4, 25.2) \\
DS75 & 3448 (12.2, 62.1, 25.7) \\
DS66 & 4211 (12.2, 60.1, 27.7) \\
DS50 & 4840 (12.5, 59.4, 28.2) \\
\hline
\end{tabular}
\end{table}

The proposed classifier model is evaluated on standard classifier evaluation measures \cite{fawcett2006classifiermeasures} such as Precision (P), Recall (R), F-measure (F) and Accuracy (A). To ensure robustness of the results, all the experiments were conducted using a stratified 10-fold cross-validation. 
\par 
The proposed method (HSC) is compared against the most commonly used domain specific dictionary (LM \cite{Loughran2011}) based method (W-Loughran), and the most recent LPS model \cite{Malo2014}. The dictionary based method is too simple to capture all intricacies of financial text. The LPS model is more complex requiring transformation of text using domain dictionary and inferencing of complex sentence structures using SVM based approach. HSC, on the other hand, is a relatively simple approach that transforms the text into a set of tags and applies rule mining on the identified tags to predict sentiment. One of the major advantages of the proposed method is that the rationale for prediction can be clearly explained as it is a rule based approach.
\par 
The rule mining requires two configurable parameters namely, minimum support (minsup) and minimum confidence (minconf). We use a default minsup and minconf value of 0.5\% and 60\% respectively. A sensitivity analysis of the parameter value changes on sentiment prediction are presented in section 4.4.
\par  
The results of our initial experiments are presented in Table~\ref{table:compareHSC}. The best F-measure and accuracy values are marked in bold.  The results reveal that the proposed method is significantly better compared to other state-of-the-art methods in all of the datasets studied. The accuracy values for the neutral classes alone are the same for both LPS and HSC models. It is to be noted that there are large number of neutral examples in the dataset (about 60\%, refer to Table~\ref{table:datasetCharacteristics}) and both the models are able to learn well from these examples and achieve a very high accuracy value of over 93\%.

\begin{table}[!htb]
\caption{Comparative evaluation of HSC against other methods}
\label{table:compareHSC}
\begin{tabular}{c|c|c|c|c!{\vrule width 1pt}c|c|c!{\vrule width .9pt}c|c|c}
\hline\noalign{\smallskip}
Dataset	& Measure & \multicolumn{3}{c}{W-Loughran} & \multicolumn{3}{c}{LPS} & \multicolumn{3}{c}{HSC} \\
& & pos & neut & neg & pos & neg & neut & pos & neg & neut \\
\noalign{\smallskip}
\hline 
\noalign{\smallskip}
DS100 & P & 0.56 & 0.64 & 0.36 & 0.74 & 0.85 & 0.84 & 0.83 & 0.93 & 0.86 \\
&	R	 & 0.13	& 0.91	& 0.16	& 0.74	& 0.87	& 0.79	& 0.82	& 0.93	& 0.81 \\
& 	F	& 0.20	& 0.75	& 0.22	& 0.74	& 0.86	& 0.81	& \textbf{0.83}	& \textbf{0.93}	& \textbf{0.83} \\
& 	A	& 0.76	& 0.63	& 0.85	& 0.87	& 0.83	& \underline{\textbf{0.95}}	& \textbf{0.91}	& \textbf{0.92}	& \underline{\textbf{0.95}} \\
DS75 &	P	& 0.60	& 0.65	& 0.38	& 0.69	& 0.83	& 0.76	& 0.77	& 0.90	& 0.83 \\
&	R	& 0.17	& 0.90	& 0.20	& 0.66	& 0.84	& 0.80	& 0.75	& 0.90	& 0.78 \\
&	F	& 0.26	& 0.76	& 0.26	& 0.67	& 0.83	& 0.78	& \textbf{0.76}	& \textbf{0.90}	& \textbf{0.80} \\
&	A	& 0.76	& 0.64	& 0.86	& 0.84	& 0.79	& \underline{\textbf{0.95}}	& \textbf{0.88}	& \textbf{0.88}	& \underline{\textbf{0.95}} \\
DS66 & P	& 0.61	& 0.63	& 0.40	& 0.60	& 0.86	& 0.73	& 0.75	& 0.86	& 0.80 \\
&	R	& 0.18	& 0.89	& 0.21	& 0.82	& 0.71	& 0.77	& 0.70	& 0.88	& 0.75 \\
&	F	& 0.28	& 0.74	& 0.28	& 0.69	& 0.77	& 0.75	& \textbf{0.72}	& \textbf{0.87}	& \textbf{0.77} \\
&	A	& 0.74	& 0.62	& 0.87	& 0.80	& 0.75	& \underline{\textbf{0.94}}	& \textbf{0.85}	& \textbf{0.84}	& \underline{\textbf{0.94}} \\
DS50	& P	& 0.57	& 0.62	& 0.41	& 0.65	& 0.76	& 0.72	& 0.72	& 0.84	& 0.79 \\
&	R	& 0.19	& 0.88	& 0.22	& 0.54	& 0.81	& 0.77	& 0.66	& 0.85	& 0.71 \\
&	F	& 0.28	& 0.73	& 0.29	& 0.59	& 0.78	& 0.74	& \textbf{0.69}	& \textbf{0.85}	& \textbf{0.75} \\
&	A	& 0.73	& 0.61	& 0.86	& 0.79	& 0.74	& \underline{\textbf{0.93}}	& \textbf{0.82}	& \textbf{0.81}	& \underline{\textbf{0.93}} \\
\hline
\noalign{\smallskip}
\end{tabular}
The best F \& A cases are marked in \textbf{bold} and the tie cases are \underline{underlined}.
\end{table}

\subsection{Study of the effect of performance indicators}
In the next set of experiments, we study the influence of performance indicators and financial sentiment words on sentiment prediction. Three different scenarios were considered for the analysis: (1) using lagging indicators (along with directionality), (2) using both lagging and leading indicators (along with directionality), and (3) using all of the tags including financial sentiment words (baseline case).
The analysis results are presented in Table~\ref{table:comparePerfIndicators}. The best F-measure and accuracy values are marked in bold. In addition, the values that are tied are underlined.
\par  
The results present a few interesting insights, when it is analysed in relation to the characteristics of the annotated datasets. First, the results for DS100 dataset using only the lagging indicators is almost close to that of the baseline case. The performance difference between the two cases widens as we navigate down the table (DS75, DS66 and DS50). Second, the performance results improve as the leading indicators are included in the model. Further performance improvements are observed when all of the tags are used. 
One can observe a correlation between inter-annotator agreement and the use of indicators. A model that uses only the lagging indicators works quite well when the inter-annotator agreement is 100\%. The second model that use both lagging and leading indicators perform better, even when the inter-annotator agreement declines to 75\%. Finally, the third model works best, even when the inter-annotator agreement declines to a very low value of 50\%. These results imply that human annotators, experienced in finance domain, have less agreement when a financial text uses leading indicators to describe company's performance. This is in line with our expectation, as the relationship amongst the leading indicators, investor sentiment and the firm’s future performance are often unclear. One can also find evidence from the finance and accounting literature on the lack of clear relationship between leading indicators and firm performance \cite{ittner1998nonfinancial}. Let us consider the following two sentences:

\begin{enumerate}[leftmargin=1cm]
\item VDW combined with LXE devices enhances productivity, enabling workers to use a single device to perform voice, scanning and keyboard functions.
\item Last year, UPM cut production, closed mills in Finland and slashed 700 jobs. 
\end{enumerate}

The first sentence refers to worker productivity improvement (a leading indicator) and the second sentence refers to cut in production and jobs (leading indicators). Both of these sentences do not have 100\% agreement and are annotated by 75\% of the annotators as polarized i.e. the first and second sentences are marked respectively as positive and negative.
\par 
The financial texts also contain general positive opinions. For example, let us consider the following sentence:
\par 
\vspace{0.5em}
"He believes that the soy-oats have a \underline{good} chance of entering the UK market"
\vspace{0.001em}
\par 
The above sentence contains neither lagging nor leading indicators but expresses a positive outlook. A high level of disagreement amongst annotators was observed and only 50\% of the annotators have marked the above sentence as positive. Another sentence that has been rated as positive by 50\% of the annotators was:  "The company’s \underline{strength} is its Apetit brand".

\begin{table}[!htb]
\caption{Influence of performance indicators on sentiment prediction}
\label{table:comparePerfIndicators}
\begin{tabular}{c|c|c|c|c!{\vrule width 1pt}c|c|c!{\vrule width 1pt}c|c|c}
\hline\noalign{\smallskip}
\scriptsize{Dataset}	& \scriptsize{Measure} & \multicolumn{3}{c}{HSC: LagInd Only} & \multicolumn{3}{c}{HSC:LagInd,LeadInd} & \multicolumn{3}{c}{HSC: All} \\
& & pos & neut & neg & pos & neg & neut & pos & neg & neut \\
\noalign{\smallskip}
\hline 
\noalign{\smallskip}
DS100	& P	& 0.89	& 0.91	& 0.87	& 0.89	& 0.92	& 0.87	& 0.83	& 0.93	& 0.86 \\
&	R	& 0.77	& 0.96	& 0.79	& 0.80	& 0.96	& 0.78	& 0.82	& 0.93	& 0.81 \\
&	F	& 0.82	& 0.93	& 0.82	& \textbf{0.84}	& \textbf{0.94}	& 0.82	& 0.83	& 0.93	& \textbf{0.83} \\
&	A	& 0.91	& 0.91	& \underline{\textbf{0.95}}	& \textbf{0.92}	& \underline{\textbf{0.92}}	& \underline{\textbf{0.95}}	& 0.91	& \underline{\textbf{0.92}}	& \underline{\textbf{0.95}} \\
DS75	& P	& 0.82	& 0.85	& 0.86	& 0.82	& 0.88	& 0.84	& 0.77	& 0.90	& 0.83 \\
&	R	& 0.65	& 0.94	& 0.68	& 0.70	& 0.94	& 0.76	& 0.75	& 0.90	& 0.78 \\
&	F	& 0.73	& 0.89	& 0.75	& \underline{\textbf{0.76}}	& \textbf{0.91}	& \underline{\textbf{0.80}}	& \underline{\textbf{0.76}}	& 0.90	& \underline{\textbf{0.80}} \\
&	A	& 0.87	& 0.86	& 0.94	& \textbf{0.89}	& \underline{\textbf{0.88}}	& \underline{\textbf{0.95}}	& 0.88	& \underline{\textbf{0.88}}	& \underline{\textbf{0.95}} \\
DS66	& P	& 0.79	& 0.80	& 0.81	& 0.79	& 0.84	& 0.81	& 0.75	& 0.86	& 0.80 \\
&	R	& 0.58	& 0.93	& 0.57	& 0.63	& 0.91	& 0.72	& 0.70	& 0.88	& 0.75 \\
&	F	& 0.67	& 0.86	& 0.67	& 0.70	& \underline{\textbf{0.87}}	& 0.76	& \textbf{0.72}	& \underline{\textbf{0.87}}	& \textbf{0.77} \\
&	A	& 0.84	& 0.81	& 0.93	& \underline{\textbf{0.85}}	& \underline{\textbf{0.84}}	& \underline{\textbf{0.94}}	& \underline{\textbf{0.85}}	& \underline{\textbf{0.84}}	& \underline{\textbf{0.94}} \\
DS50	& P	& 0.75	& 0.77	& 0.82	& 0.75	& 0.81	& 0.80	& 0.72	& 0.84	& 0.79 \\
&	R	& 0.54	& 0.91	& 0.51	& 0.59	& 0.90	& 0.69	& 0.66	& 0.85	& 0.71 \\
&	F	& 0.62	& 0.83	& 0.63	& 0.66	& \underline{\textbf{0.85}}	& 0.74	& \textbf{0.69}	& \underline{\textbf{0.85}}	& \textbf{0.75} \\
&	A	& 0.82	& 0.78	& 0.92	& \textbf{0.83}	& 0.80	& \underline{\textbf{0.93}}	& 0.82	& \textbf{0.81}	& \underline{\textbf{0.93}} \\
\hline
\end{tabular}
\end{table}

From the foregoing discussions and the experimental results, it is clear that financial texts are interpreted differently based on the nature of the indicators and sentiment words. It is to be noted that sentiment analysis in other domains (such as movies, music) utilizes general sentiment lexicon words to predict sentiments. In the financial sentiment analysis literature, similar approach has been borrowed with a refinement of dictionary words to finance domain. This paper presents a new perspective and suggests the use of multiple levels of analysis (lagging indicator, leading indicator, domain specific lexicon words) to improve the quality of financial sentiment analysis. The experimental results clearly demonstrate the utility of such a multi-level sentiment analysis. 
\par 
The multi-level approach to financial sentiment analysis can be very useful in building models to suit specific application requirements. For example, an investor might prefer a model built using lagging (or lagging and leading) indicators over a more accurate baseline model.  The former model is more likely to help the investor make the best investment decisions, even though it offers relatively lower accuracy.

\subsection{Analysis of alternate classifier models}
In the financial sentiment analysis context, the number of examples in each of the classes is likely to be highly unbalanced. Therefore, a hierarchical sentiment classifier was used in this paper. Table~\ref{table:compareClassifiers} presents an analysis of hierarchical versus multi-class and one-against-one versions of the associative classifier. A comparison of multi-class versus hierarchical classifier reveals that the multi-class version performs marginally better for DS100 dataset. However, the performance degrades considerably for other datasets. For example, the F-measure values for DS50 dataset is 64\% and 69\% respectively for multi-class and hierarchical versions. 
\par 
A comparison of one-against-one versus hierarchical classifier suggests that the hierarchical classifier performs better. However, the margin of difference is not very significant, except for a few cases (e.g. D75 dataset, negative class values of 74\% versus 80\%). Overall, the comparative evaluation of different variants of the classifier models did not indicate statistically significant differences in the observed results. Future work could consider detailed evaluation of alternate classifier models on more diverse datasets to better understand the trade-offs involved.

\begin{table}[!htb]
\caption{Performance analysis of One-against-One, Multi-class and HSC classifiers}
\label{table:compareClassifiers}
\begin{tabular}{c|c|c|c|c!{\vrule width 1pt}c|c|c!{\vrule width .9pt}c|c|c}
\hline\noalign{\smallskip}
\scriptsize{Dataset}	& \scriptsize{Measure} & \multicolumn{3}{c}{One-against-one} & \multicolumn{3}{c}{Multi-class} & \multicolumn{3}{c}{HSC} \\
& & pos & neut & neg & pos & neg & neut & pos & neg & neut \\
\noalign{\smallskip}
\hline 
\noalign{\smallskip}
DS100	& P	& 0.84	& 0.92	& 0.87	& 0.85	& 0.94	& 0.84	& 0.83	& 0.93	& 0.86 \\
&	R	& 0.82	& 0.94	& 0.76	& 0.81	& 0.92	& 0.89	& 0.82	& 0.93	& 0.81 \\
&	F	& \underline{\textbf{0.83}} & \underline{\textbf{0.93}}	& 0.81	& \underline{\textbf{0.83}}	& \underline{\textbf{0.93}}	& \textbf{0.86}	& \underline{\textbf{0.83}}	& \underline{\textbf{0.93}}	& 0.83 \\
&	A	& \underline{\textbf{0.91}}	& 0.91	& 0.94	& \underline{\textbf{0.91}}	& \underline{\textbf{0.92}}	& \underline{\textbf{0.95}}	& \underline{\textbf{0.91}}	& \underline{\textbf{0.92}}	& \underline{\textbf{0.95}} \\
DS75	& P	& 0.78	& 0.88	& 0.85	& 0.78	& 0.89	& 0.83	& 0.77	& 0.90	& 0.83 \\
&	R	& 0.75	& 0.92	& 0.66	& 0.71	& 0.90	& 0.81	& 0.75	& 0.90	& 0.78 \\
&	F	& \underline{\textbf{0.76}}	& 0.89	& 0.74	& 0.74	& 0.89	& \textbf{0.82}	& \underline{\textbf{0.76}}	& \textbf{0.90}	& 0.80 \\
&	A	& \textbf{0.88}	& 0.86	& 0.94	& 0.87	& 0.87	& \underline{\textbf{0.95}}	& 0.87	& \textbf{0.88}	& \underline{\textbf{0.95}} \\
DS66	& P	& 0.76	& 0.85	& 0.81	& 0.75	& 0.84	& 0.80	& 0.75	& 0.86	& 0.80 \\
&	R	& 0.69	& 0.89	& 0.70	& 0.64	& 0.88	& 0.76	& 0.70	& 0.88	& 0.75 \\
&	F	& \textbf{0.73}	& \underline{\textbf{0.87}}	& 0.75	& 0.69	& 0.86	& \textbf{0.78}	& 0.72	& \underline{\textbf{0.87}}	& 0.77 \\
&	A	& \underline{\textbf{0.85}}	& 0.83	& \underline{\textbf{0.94}}	& 0.84	& 0.82	& \underline{\textbf{0.94}}	& \underline{\textbf{0.85}}	& \textbf{0.84}	& \underline{\textbf{0.94}} \\
DS50	& P	& 0.73	& 0.82	& 0.80	& 0.78 & 0.80	& 0.79	& 0.72	& 0.84	& 0.79 \\
& 	R	& 0.65	& 0.87	& 0.67	& 0.54	& 0.89	& 0.73 & 0.66	& 0.85	& 0.71 \\
&	F	& \underline{\textbf{0.69}}	& 0.84	& 0.73	& 0.64	& 0.84	& \textbf{0.76}	& \underline{\textbf{0.69}}	& \textbf{0.85}	& 0.75 \\
&	A	& \textbf{0.83}	& 0.80	& 0.93	& 0.81	& 0.79	& \textbf{0.94}	& 0.82	& \textbf{0.81}	& 0.93 \\
\hline
\end{tabular}
\end{table}

\subsection{Comparative evaluation of HSC and machine learning models}
In the recent literature on financial sentiment analysis, machine learning approaches like Naive Bayes (NB) and Support Vector Machines (SVM) have been explored. These approaches utilize bag-of-words or n-grams or sentiment words as features for building the model. Such models generate large number of features and pose scalability issues. Besides, the models based on bag-of-words (or n-grams) are shown to be less predictive in the past research studies \cite{li2014news,schumaker2009textual}. 
\par 
We analyze the use of machine learning models with parsimonious set of features (tags) introduced in this paper.  A term document matrix was built using the tags as features. The generated term document matrix was then weighted with the help of standard Term frequency-Inverse document frequency (TF-IDF) measure. The standard python library (sklearn) was utilized for building NB and SVM models. For SVM, the default parameter values of C, $\gamma$ and RBF kernel were used for the experimentation.   
\par
\begin{figure}[tbh]
\centering
\includegraphics[width=11.9cm,height=6cm]{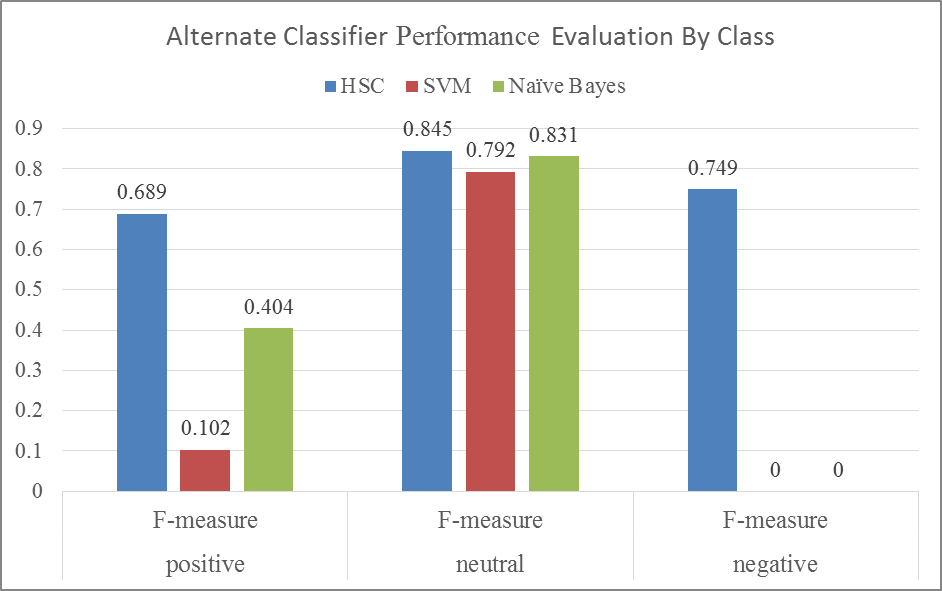}
\caption{Classifier Performance for DS50 dataset}
\label{figure:alternateClassifierPerf}
\end{figure}

 
The results of our experiments are shown in Fig.~\ref{figure:alternateClassifierPerf}. The results reveal that HSC perform better compared to other models. SVM and NB models produced a zero F-measure values for the negative class. This can be attributed to the models inability to learn from the parsimonious set of features. The naive bayes method was found to perform better than SVM for the positive and neutral classes. 
It is evident from the results that the standard machine learning models produced poor results compared to HSC. These results clearly demonstrate that an associative classifier (HSC) is quite effective in predicting sentiments even with parsimonious set of features.

\subsection{Effect of changes in rule mining parameters}

\begin{figure}[tbh]
\centering
\includegraphics[width=11.9cm,height=8cm]{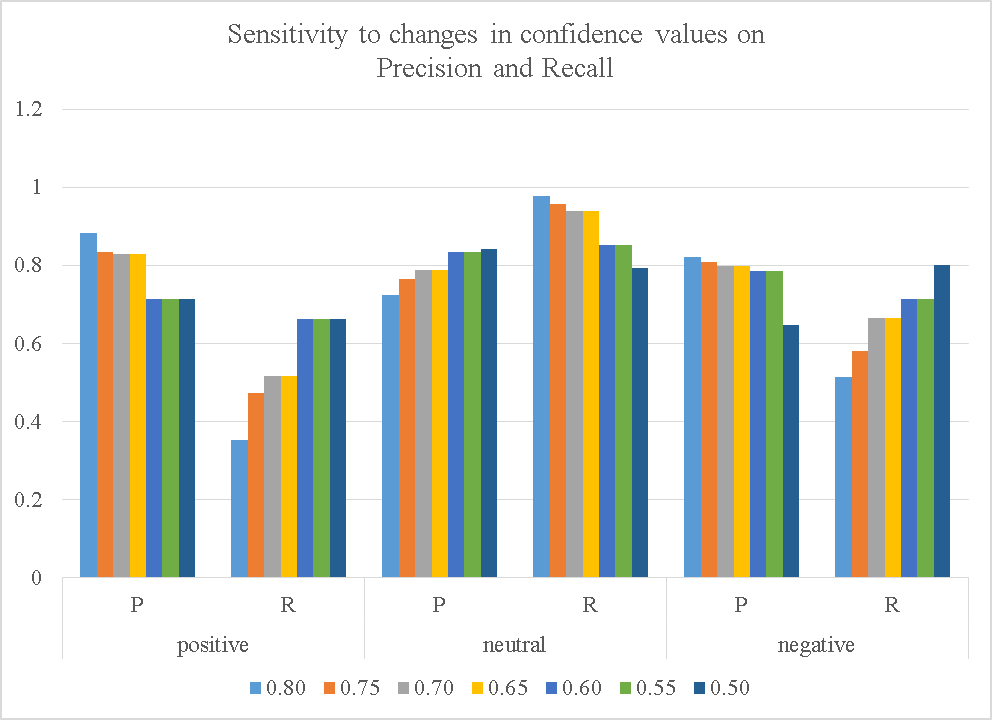}
\caption{Performance analysis of confidence parameter changes for DS50 dataset}
\label{figure:sensitivityAnalysis}
\end{figure}

The association rule mining requires two parameters to be specified, namely, minimum support and minimum confidence. The minimum confidence parameter influences the quality of rules generated and plays a critical role in sentiment prediction. We study the influence of the choice of minimum confidence values on the precision and recall values of the classifier. The results of our investigation are presented in Fig.~\ref{figure:sensitivityAnalysis}.
\par 
The results show the trade-off involved between precision and recall. An increase in confidence value is associated with increase in precision for positive and negative classes. But, the precision increase at higher confidence values comes at the cost of lower recall. The observed relationship is quite intuitive and is in line with our expectations. One can choose an appropriate value of confidence based on the level of precision and recall desired. Alternately, one can choose an optimal threshold value that maximizes F-measure. It is to be noted that F-measure is the harmonic mean of precision and recall. 

\subsection{Study of the use of performance indicator - directionality reversal}
In the next set of experiments, we analyze the influence of directionality reversals. Let us consider the following two sentences to understand the need for directionality reversals. 

\begin{enumerate}[leftmargin=1cm]
\item	\textit{Turnover} \underline{rose} to EUR 21mn from EUR 17 mn. 
\item Unit \textit{costs} for flight operations \underline{fell} by 6.4 percent.
\end{enumerate}

Both of the above sentences belong to the positive class. The above two sentences will be tagged respectively as LagInd::UP and LagInd::DOWN. It is reasonable to assume that one is likely to encounter more number of LagInd::UP cases in positive class and more number of LagInd::DOWN cases in negative class. Hence, an associative classifier (HSC) is most likely to predict the class for the second example as negative. 
\par 
In the literature, Malo et al \cite{Malo2014} use a lexicon to pre-define directional dependence for each financial term. The baseline associative classifier presented in this paper does not use such directional dependencies. In order to understand the influence of directional dependencies on classifier performance, a directionality reversal is applied as a post-processing operation. That is, after the initial tagging, a lookup is made for specific indicators (like operating cost, operating loss, expenses) and the directionality is reversed. For the illustrative example in sentence 2 above, the tag is reversed from LagInd::DOWN to LagInd::UP as lower cost is considered as a positive outcome.  

\begin{table}[!htb]
\caption{Experimental analysis of PI and Directionality reversal}
\label{table:directionalityAnalysis}
\begin{tabular}{c|c|c|c|c!{\vrule width 1pt}c|c|c}
\hline\noalign{\smallskip}
Dataset	& Measure & \multicolumn{3}{c}{HSC} & \multicolumn{3}{c}{PI-Directionality reversal} \\
& & pos & neut & neg & pos & neg & neut \\
\noalign{\smallskip}
\hline 
\noalign{\smallskip}
DS100	& P	& 0.83	& 0.93	& 0.86	& 0.83	& 0.93	& 0.89 \\
&	R	& 0.82	& 0.93	& 0.81	& 0.85	& 0.93	& 0.80 \\
&	F	& 0.83	& \underline{\textbf{0.93}} & 0.83	& \textbf{0.84}	& \underline{\textbf{0.93}}	& \textbf{0.84} \\
&	A	& 0.91	& \underline{\textbf{0.92}}	& 0.95	& \textbf{0.92} & \underline{\textbf{0.92}}	& \textbf{0.96} \\
DS75	& P	& 0.77	& 0.90	& 0.83	& 0.77	& 0.90	& 0.85 \\
&	R	& 0.75	& 0.90	& 0.78	& 0.77	& 0.90	& 0.77 \\
&	F	& 0.76	& \underline{\textbf{0.90}}	& 0.80	& \textbf{0.77}	& \underline{\textbf{0.90}}	& \textbf{0.81} \\
&	A	& \underline{\textbf{0.88}}	& \underline{\textbf{0.88}}	& \underline{\textbf{0.95}}	& \underline{\textbf{0.88}}	& \underline{\textbf{0.88}}	& \underline{\textbf{0.95}} \\
DS66	& P	& 0.75	& 0.86	& 0.80	& 0.75	& 0.86	& 0.82 \\
&	R	& 0.70	& 0.88	& 0.75	& 0.71	& 0.88	& 0.74 \\
&	F	& 0.72	& \underline{\textbf{0.87}}	& 0.77	& \textbf{0.73}	& \underline{\textbf{0.87}}	& \textbf{0.78}\\
&	A	& \underline{\textbf{0.85}}	& \underline{\textbf{0.84}}	& 0.94	& \underline{\textbf{0.85}}	& \underline{\textbf{0.84}}	& \textbf{0.95} \\
DS50	& P	& 0.72	& 0.84	& 0.79	& 0.71	& 0.84	& 0.81 \\
&	R	& 0.66	& 0.85	& 0.71	& 0.68	& 0.85	& 0.71 \\
&	F	& 0.69	& \underline{\textbf{0.85}}	& \underline{\textbf{0.75}}	& \textbf{0.70}	& \underline{\textbf{0.85}}	& \underline{\textbf{0.75}} \\
&	A	& 0.82	& \underline{\textbf{0.81}}	& 0.93	& \textbf{0.83}	& \underline{\textbf{0.81}}	& \textbf{0.94} \\
\hline
\end{tabular}
\end{table}

The results of the directionality reversal experiments are presented in Table~\ref{table:directionalityAnalysis}. The results show that there is a general improvement in classifier performance for both positive and negative classes. This is in line with our expectation as the positive and negative example cases are correctly tagged with directionality reversals. The observed improvement is marginally better compared to the baseline case but the results are statistically insignificant. These results demonstrate that the proposed model (HSC) offers better predictive results even when the directionality reversal words are not used. 
\par 
An alternate way to handle the directional dependence is to tag the identified indicators along with the business terms. For example, while tagging sentence 'Operational cost rose by 30\%', the term 'operational cost' can be tagged along with 'LagInd::UP'. The association rule mining can then be applied on the transactions that contain both performance indicators and the associated terms. The main challenge with this approach is that the support of individual terms (like cost, expenses) will be very low and is unlikely to satisfy the minimum support constraints or produce models with very poor accuracy (if very low minimum support value is used). In such cases, the use of more complex variable support rule mining approaches can be explored. 

\subsection{Discussion}
This paper proposed a hierarchical sentiment classifier using performance indicators for financial sentiment prediction. The proposed approach was found to be quite effective achieving an accuracy of over 80\% and F-measure of over 68\% in all of the datasets studied. 
\par 
The findings of this paper have implications for both theory and practice. From a theoretical perspective, the paper presents a novel approach to financial sentiment analysis using performance indicators. The presented approach is more closely aligned with the way humans interpret financial texts while making investment decisions. The associative sentiment classifier generates easily explainable predictions compared to other methods available in the literature. Our rigorous experimental evaluation clearly demonstrate the utility of the approach over the state-of-the-art methods. Furthermore, the results observed in Table~\ref{table:comparePerfIndicators} clearly reveals the reasons for differences amongst annotators with the help of performance indicators (lagging and leading indicators).    
\par 
The paper presents new perspectives on the influence of performance indicators and financial sentiment lexicon words on sentiment prediction. The experimental results reveal that it is useful to consider financial sentiment analysis at multiple levels (using lagging indicator, combination of lagging and leading indicators, and all indicators along with sentiment words). An investor, analyst or financial institution can benefit by choosing an appropriate financial sentiment model to suit their specific application requirements e.g. track overall market outlook, make investment decision and so on.
\par 
The transparency and replicability of results is often a challenge in the financial sentiment analysis literature \cite{Loughran2016} due to the proprietary nature of datasets, dictionaries and advanced sentence parsing rules. To ensure transparency and replicability of the results presented, the detailed grammar and sentence parsing steps used for tagging are provided in Appendix. 

\section{Conclusion}
This paper examined the use of performance indicators for predicting sentiments from financial texts. It presented a hierarchical classifier that uses the concept of association rules. The effectiveness of the classifier was demonstrated through rigorous experimental analysis on a benchmark financial dataset. 
\par 
A study of the influence of performance indicators on sentiment prediction revealed interesting insights. The presence of varying levels of influence of lagging indicators, leading indicators and sentiment words on sentiment prediction were observed for different datasets. The results are clearly in alignment with the way humans interpret financial texts and make decisions.     
\par 
As part of our future work, we plan to explore several interesting extensions. First, a more fine-grained analysis of performance indicators is likely to reveal interesting insights. For instance, a balanced scorecard approach \cite{Kaplan1996} analyses company's performance from multiple dimensions, namely, financial, customers, internal business processes and learning \& growth. Performance indicators can be categorized across each of these dimensions and the influence of each of the dimensions on financial sentiment can be explored. Second, advanced variable support rule mining methods can be studied to capture the directional dependencies in financial texts without using a lexicon. Third, the proposed method is not tuned to handle sentences that have mix of positive and negative orientations on performance indicators. One can investigate the use of utility mining \cite{liu2012mining} approaches to give different weights to different indicators. Such an approach allows one to capture varying influence of indicators and further improve sentiment prediction.  
\par 
The findings of this study are likely to be useful for financial institutions in building superior sentiment analysis models. The investors or financial analysts can make better investment decisions using qualitative financial texts.

\bibliography{biblio} 

\newpage

\textbf{Appendix}

\vspace{0.1cm}
\par 
\textbf{A. Parsing sentences using the NLTK toolkit} \\

\begin{figure}[h]
\hspace{5mm}
\begin{minipage}{40em}
\vspace{0.5em}
\textbf{Grammar} \\

  $JJ: \{<JJ.*>*\} \\
  VB: \{<VB.*>\} \\
  NP: \{(<NNS|NN>)*\} \\
  NPP: \{<NNP|NNPS>\} \\
  RB: \{ <RB.*> \} \\
  NPJJ: \{ \\
    \hspace*{1.5em} (((<NP|NPP>+<IN><.*>*<,>)|(<JJ>*<NP|NPP>+ \\
\hspace*{2.5em} <VB><NPP>*<.*>*<,>)) \\
\hspace*{2.5em} (<RB> | <DT><NP> | <VB><TO>) )      |  \\
\hspace*{1.5em}     <JJ>*<NP>(<IN><DT>*<JJ>*<NPP>*<NP>*)*<VB>|      \\
\hspace*{1.5em}     <JJ>+<NP><VB>      |  \\
\hspace*{1.5em}     <NP|NPP>+(<(><.*>*<)>)*((<IN><JJ>*<NP>) \\
\hspace*{2.5em} (<IN><CD>)*)*<VB>+           | \\
\hspace*{1.5em}     <NP><.*>+(<RB>|<JJ>)      | \\
\hspace*{1.5em}     <NP|NPP>+(<IN><NP|NPP>)*<.*>*<VB> \\
\hspace*{2.5em}(<DT><JJ>)*      | \\
\hspace*{1.5em}     <NP><VB>(<RB>|(<TO><DT><NP>)) |       \\
\hspace*{1.5em}     <VB><NP|NPP>*<POS><JJ>*<NP>       | \\
\hspace*{1.5em}     <VB><PRP.*><JJ>*<NP><IN> |    \\
\hspace*{1.5em}     <VB><TO>*<DT>*<JJ>*<NP><IN>*<NP>*     | \\
\hspace*{1.5em}     <NP|NPP>+(<IN><DT>*<RB>*<JJ>*<NP|NPP>)* \\
\hspace*{2.5em} <RB>*(<VB><JJ><NP>)*<VB> \\
\hspace*{2.5em}(<DT><CD><NP>)*      | \\
\hspace*{1.5em}     (<JJ>)*<NP|NPP>+<.*>*(<,><.*>*<,>)*<NP> \\
  \}$
\end{minipage}
\end{figure}

\underline{Steps:}
\begin{enumerate}[leftmargin=1cm]
\item Parse the sentence with a regular expression parser using the grammar specified above

\item If a match for 'NPJJ' tree pattern is found:
\begin{enumerate}[label=(\alph*)]
\item Traverse the subtree to get the combination of NP (potential performance indicator) and JJ/RB/VB (potential directionality word). Look up the dictionary for the matching indicator and directionality word.
\item If the combination of NP and JJ/RB/VB are not found, check for presence of individual words (either performance indicator or directionality word) in the dictionary
\item Tag the sentence based on the identified matches.
\end{enumerate}

\end{enumerate}

\vspace{.5cm} 
\par 
\textbf{B.	Parsing numeric values to determine directionality}
\vspace{0.3cm}
\par 
\underline{Preconditions:}
\begin{enumerate}[leftmargin=1cm]
\item If a sentence has not been tagged with combination of performance indicators and directionality using the parse rules specified in Section A above. 
\item The sentence contains terms like compared to, versus, down from, up from  \textit{\#common sentences where one is likely to observe multiple numeric values without the use of directional words.}
\end{enumerate}

\vspace{0.3cm}
\par 
\underline{Example sentence:} \textit{Operating profit} margin was 8.3\%, compared to 11.8\% \\
\hspace*{3.5cm}  a year earlier. \\
\par 
\underline{Expected tag output:} LagInd::DOWN \\
\par 

\begin{figure}[h]
\hspace{5mm}
\begin{minipage}{40em}
\vspace{0.5em}
\textbf{Grammar} \\

  $JJ: \{<JJ.*>*\} \\
  VB: \{<VB.*>\} \\
  NP: \{(<NNS | NN>)*\} \\
  NPP: \{<NNP|NNPS>\} \\
  RB: \{ <RB.*> \} \\
  CD: \{ <CD> \} \\
  NPJJ: \{$  \\
     \hspace*{1.5em} $<NP|NPP>+(<(><.*>*<)>)*(<IN><DT>*<RB>* \\
     \hspace*{3em} <JJ>*<NP|NPP>)* \\
     \hspace*{3em}         <RB>*(<VB><JJ><NP>)*<VB>(<DT><CD><NP>)* \\
     \hspace*{3em}         <NP|NPP>*<CD>*<.*>*<CD>* |        \\
     \hspace*{1.5em}<NP|NPP><IN><NP|NPP><CD><.*>*<,><VB> \\
     \hspace*{3em}         <IN><NP|NPP><CD> $ \\
    \} 
\end{minipage}
\end{figure}

\par 
\underline{Steps:}
\begin{enumerate}[leftmargin=1cm]
\item Parse the sentence with a regular expression parser using the grammar specified above

\item If a match for 'NPJJ' tree pattern is found:
\begin{enumerate}[label=(\alph*)]
\item Traverse the subtree to get the combination of NP, CD, CD to extract the performance indicator and numeric values. The numeric values are analysed to determine the directionality (UP/DOWN)
\item Tag the sentence based on the identified matches.
\end{enumerate}
\end{enumerate}

\end{document}